\documentclass{ws-procs9x6}

\begin{document}

\title{The Importance of Parity-Dependence of the Nuclear Level Density
in the Prediction of Astrophysical Reaction Rates}

\author{D. MOCELJ, T. RAUSCHER, G. MART\'INEZ-PINEDO}

\address{Departement f\"ur Physik und Astronomie\\
Universit\"at Basel\\
Basel, Switzerland\\
E-mail: Thomas.Rauscher@unibas.ch}

\author{Y. ALHASSID}

\address{Center for Theoretical Physics\\ 
Sloane Physics Laboratory, Yale University\\
New Haven, CT, USA}


\maketitle

\abstracts{
A simple description for obtaining the parity distribution of nuclear
levels in the {\it pf + g$_{9/2}$} shell
as a function of excitation energy was recently derived. We implement
this in a global nuclear level density model. In the framework of the
statistical model, cross sections and astrophysical reaction rates are
calculated in the Fe region and compared to rates obtained with the
common assumption of an equal distribution of parities. We find
considerable differences, especially for reactions involving particles
in the exit channel.
}

\section{Introduction}
The nuclear level density is an important ingredient in the prediction 
of nuclear reaction rates in astrophysics. So far, all theoretical,
global calculations of astrophysical rates assume an equal
distribution of the state parities at all energies. It is obvious that
this assumption is not valid at low excitation energies of a
nucleus. However, a globally applicable recipe was lacking. We combine
a formula for the energy-dependent parity distribution with a
microscopic-macroscopic nuclear level density\cite{rau}. The formula
describes well the transition from low excitation energies where a
single parity dominates to high excitations where the two densities
are equal. It was tested against Monte Carlo shell model
calculations. This treatment is further applied in the calculation of
astrophysical reaction rates for nuclei in the Fe region. 
\section{Method}
Y. Alhassid {\it et.\ al}\cite{alha} have calculated the ratio
$Z_-/Z_+$ for nuclei in the iron region using the complete {\it pf +
  g$_{9/2}$} shell. In
Fig.~1 the ratio of odd to even parity states as a function of inverse
temperature $\beta$ for three nuclei in the iron region is shown. 
\begin{figure}[t!]
\centerline{\epsfxsize=3.9in\epsfbox{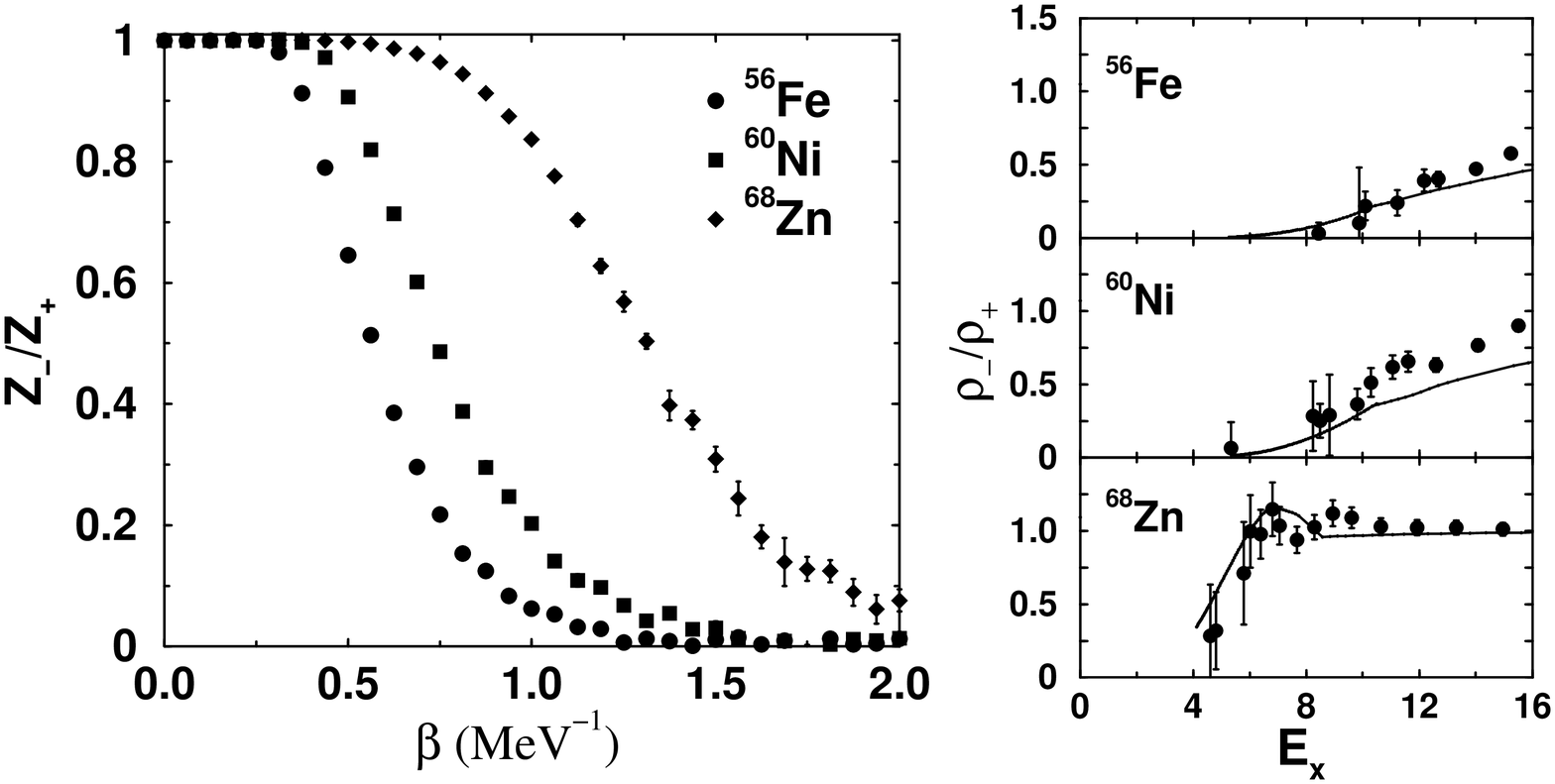}}   
\caption{{\it Left}: Ratio of odd to even parity partition function
  $Z_-/Z_+$ versus inverse temperature $\beta$ for $^{56}$Fe, $^{60}$Ni and
 $^{68}$Zn, calculated with the Monte Carlo
  Method\protect\cite{alha}.{\it Right}: The parity ratio
  $\rho_-/\rho_+$  versus
  excitation energy $E_x$. The solid lines are calculated using
  Eq. (1), and the solid circles are obtained with the Monte Carlo
  Method of Nakada {\it et.\ al}\protect\cite{nak}. }
\end{figure}
The observed parity dependence can be explained quantitively by a
simple model. Single particle levels are divided into two groups,
according to their individual parities. The group which has the smaller
average occupation number is denoted by $\Pi$. The distribution of the
occupancies of the $\Pi$ parity group can be assumed to be Poisson, if
the single particle states are occupied independently and randomly:
$$P(n)=\frac{f^n}{n!}e^{-f} \ .$$ Here $f$ is the average occupancy of
orbitals with parity $\Pi$. The probability to have an odd/even parity
state is given by: 
\begin{eqnarray*}
P_+(n) &=& \sum_{n,even}P(n)=e^{-f}\cosh f \\
P_-(n) &=& \sum_{n,odd} P(n)=e^{-f}\sinh f \ , 
\end{eqnarray*}
and their ratio by 
\begin{equation} 
\frac{P_-}{P_+}=\frac{Z_-}{Z_+}=\tanh f \ . 
\end{equation}
The arguments leading to this equation are easily extended to the case
where the protons and neutrons are treated separately --- $f$ has only
to be replaced by the sum of individual contributions from neutrons
and protons. 

The back-shifted Fermi-gas model (BBF) of Rauscher {\it et.\
  al}\cite{rau} is used to calculate the total partition function $Z$
  at constant excitation energy. Using $Z_+ + Z_- = Z$ and Eq. (1),
we can determine $Z_{+/-}$ and calculate the thermal energies for
even- and odd parity states. Canonical entropies, heat capacities and
the parity projected level densities are calculated from standard
thermodynamic relations. 
\section{Results and Discussion}
We have used the parity dependent level density to calculate
astrophysical reaction rates in the global Hauser-Feshbach (HF) model
{\it NON-SMOKER}\cite{raua} and compared the results to the standard
values\cite{raub}. We show the parity dependence for three nuclides,
$^{64}$Fe,$^{66}$Ni, $^{68}$Zn and investigated all reaction channels
involving these nuclei. 
\begin{figure}[t!]
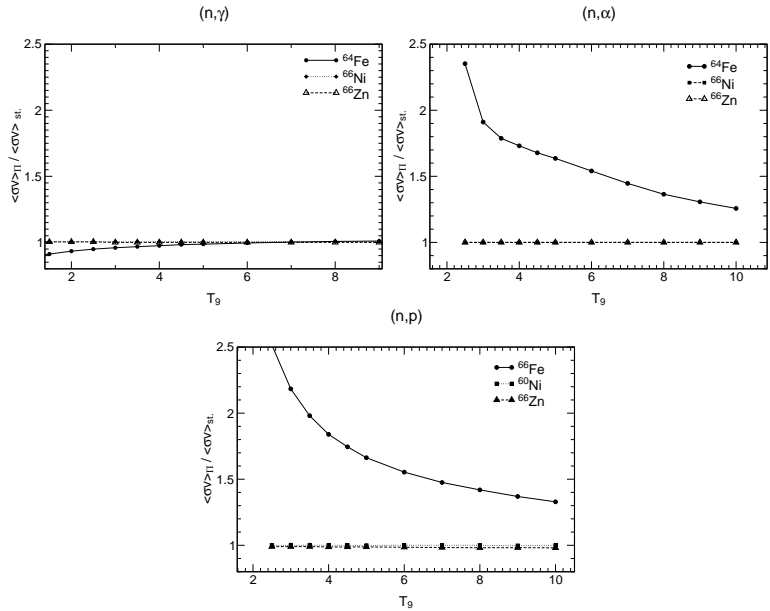

\begin{center}
\epsfxsize=5cm   %width of figure - will enlarge/reduce the figures
\epsfbox{prag_ngamma.eps}
\epsfxsize=5cm 
\epsfbox{prag_nalpha.eps}
%\figurebox{2cm}{3cm}{} %to have a box alone 
\centerline{\epsfxsize=5cm\epsfbox{prag_np.eps}}   
\end{center}
\caption{Comparison of the parity dependent reaction rates to the
  standard\protect\cite{raub}, which assumes an equal distribution of
  odd and even parity states.}
\end{figure}
The impact on the rates involving the Ni and
Zn nuclei is small and negligible compared to the remaining
uncertainties in the global HF model. This is due to the fact that a
sufficiently large number of excited states is known
experimentally. Up to 20 experimental states are considered in the
standard calculation and only above the last known state, the
theoretical level density is in effect. However, the case is different
for reactions involving $^{64}$Fe. No information on experimental
states is known here and therefore the full impact of the parity
dependence can be seen. In the $(n,\gamma)$ case $20\%$ difference
are found. Much larger differences are seen in the reactions involving
$^{64}$Fe in the final particle channel. Because of lack of negative
parities at low excitation energies, the particle emission channel
becomes strongly enhanced in all such reactions with low or negative
Q values. 
\section{Outlook}
The case for $^{64}$Fe shows that a large effect of the parity
dependence can be expected far from stability where no experimental
information on excited states is available and that it is extremely
important to include such a modified level density. The current approach
is valid only for even-even nuclei in the {\it pf + g$_{9/2}$}
shell. Work is in progress to extend this description to be able to
calculate the parity distribution for a large number of nuclei far
from stability on the proton-rich as well as neutron-rich side. 
\section*{Acknowledgments}
D.M. would like to thank K.-H. Langanke for useful advice.   
This work was supported by the Swiss NSF grants 2124-055833.98,
2000-061822.0 and 2024-067428.01.


\begin{thebibliography}{0}
\bibitem{rau} T. Rauscher, F.-K. Thielemann and K.-L. Kratz, {\it
  Phys. Rev. }{\bf C56}, 1613 (1997)
\bibitem{alha} Y. Alhassid, G. F. Bertsch, S. Liu and H. Nakada, {\it 
Phys. Rev. Lett.}{\bf 84}, 4313 (2000)
\bibitem{nak} H. Nakada and Y. Alhassid, {\it Phys. Lett.} {\bf B79},
  231 (1998)
\bibitem{raua} T. Rauscher and F.-K. Thielemann, {\it At. Data
  Nucl. Data Tables }{\bf 75}, 1 (2000)
\bibitem{raub} T. Rauscher and F.-K. Thielemann, {\it At. Data
 Nucl. Data Tables }{\bf 79}, 47 (2001)
\end{thebibliography}
\end{document}